\documentclass{aa}
\usepackage{graphicx}
\usepackage{txfonts}

\begin{document}

\title{Physical parameters and multiplicity of five southern close
eclipsing binaries\thanks{Based on observations made at the Siding
Spring Observatory, Australia.}}

\author{
T. Szalai
\inst{1} 
\and
L.L. Kiss
\inst{2,3} 
\and
Sz. M\'esz\'aros
\inst{1,4} 
\and
J. Vink\'o
\inst{1} 
\and
Sz. Csizmadia
\inst{5}
}

\offprints{L.L. Kiss, e-mail: l.kiss@physics.usyd.edu.au}

\institute{
Department of Optics and Quantum Electronics, University of
Szeged, Hungary
\and
School of Physics A28, University of Sydney, NSW 2006, Australia
\and
Department of Experimental Physics, University of Szeged, Hungary
\and
Harvard-Smithsonian Center for Astrophysics, Cambridge, MA, USA
\and
Konkoly Observatory of the Hungarian Academy of Sciences, Budapest, Hungary}

\date{Received ..; accepted ..}

\abstract{}{}{}{}{}

\abstract
{}
{Detect tertiary components of close binaries from spectroscopy and light curve 
modelling; investigate light-travel time effect and the possibility of magnetic 
activity cycles; measure mass-ratios for unstudied systems and derive absolute
parameters.}
{We carried out new photometric and spectroscopic observations of five bright 
($\langle V \rangle<10.5$ mag) close eclipsing binaries, predominantly in
the southern skies. We obtained full Johnson $BV$ light curves, which were modelled 
with the Wilson-Devinney code. Radial velocities were measured with the 
cross-correlation method using IAU radial velocity standards as spectral templates.
Period changes were studied with the O$-$C method, utilising published
epochs of minimum light (XY~Leo) and ASAS photometry (VZ~Lib). 
}
{
For three objects (DX~Tuc, QY~Hya, V870~Ara), absolute parameters have been
determined for the first time. We detect spectroscopically the tertiary 
components in XY~Leo, VZ~Lib and discover one in QY~Hya. 
For XY~Leo we update the light-time effect parameters and detect a secondary 
periodicity of about 5100 d in the O$-$C diagram that may hint about the 
existence of short-period magnetic cycles. A combination of recent photometric 
data shows that the orbital period of the tertiary star in VZ~Lib is 
likely to be over 1500 d. QY~Hya is a semi-detached 
X-ray active binary in a triple system  with K and M-type components, while
V870~Ara is a contact binary
with the third smallest spectroscopic mass-ratio for a W~UMa star to date
($q=0.082\pm0.030$). This small mass-ratio, being close to the theoretical
minimum for contact binaries, suggests that V870~Ara has the potential
of constraining evolutionary scenarios of binary mergers. 
The  inferred distances to these systems are compatible with the Hipparcos parallaxes.}
{}


\maketitle

\section{Introduction}

Modelling variations of close eclipsing binaries is a powerful method for  determining
fundamental physical parameters of stars. This is because the observed brightness, colour
and radial velocity changes give strong constraints on the  geometric configuration of a
given system. W~UMa-type stars are low-mass eclipsing contact binaries with orbital
periods  less than 1 day, showing continuous light variations. The components fill their
Roche lobes so that the strongly distorted stars touch each other at the inner Lagrangian
point. These systems have long been known as the most numerous of all stars, with roughly
one W~UMa binary per 500 ordinary dwarf stars (Rucinski 2002a), which is one of the
reasons why many of them, including bright southern ones, are left unstudied until now.
In particular, measuring radial velocities has been a tedious task with very slow
progress. For instance, Bilir et al. (2005) noted that among the 751 recorded W~UMa
binaries in the recently revised fourth edition of the General Catalog of Variable Stars,
only 129 systems were found to have radial velocities, mostly northern ones. Very
recently, Rucinski  \& Duerbeck (2006) and Duerbeck \& Rucinski (2007) published radial
velocities  for 23 predominantly southern close binaries, but apart from that,  there is
basically no dedicated program for obtaining new data in the south. There are several
important problems related to the formation, internal structure and evolution of these
systems, like the kinematics (e.g. Bilir et al. 2005), energy transfer between the
components (Csizmadia \& Klagyivik 2004 and references therein) magnetic activity  and
its cyclic nature (Borkovits et al. 2005) and the frequency of additional components of
W~UMa stars (Pribulla \& Rucinski 2006 and references therein), all in need for more
extensive data. The latter one is particularly important, because it is the angular
momentum transfer in hierarchic triple systems that can lead to such close binary systems
as W~UMa-type stars, so one can even hypothesize that {\it all} contact binaries reside
in triple or multiple systems (Pribulla \& Rucinski 2006; D'Angelo, van Kerkwijk \&
Rucinski 2006).  Testing this hypothesis is important, because angular momentum
evolution, in which magnetic braking was shown to be a key factor to explain the observed properties of contact binaries 
(Mochnacki 1981; Stepien 1995), can be strongly affected by the presence of a third body (Eggleton \&
Kiseleva-Eggleton 2001). In a wide range of
initial conditions an originally detached binary can reach the contact phase and this
process can be altered by a tertiary star in the system (for a recent review
see Eggleton 2006).

Late-type contact binary stars are known to be very active objects
with stellar spots (Maceroni \& van't Veer 1996) and X-ray radiation
due to chromospheric activity (Stepien et al. 2001; Chen et al. 2006; Geske
et al. 2006). These pieces of evidence can be interpreted by the Applegate-mechanism
(Applegate 1992), which involves the interchange of magnetic and kinetic
energy. An observable manifestation could be cyclic orbital period variation,
which seems to be present in many contact binaries, although not necessarily 
caused by magnetic activity -- for instance, light-travel time effect due to a third body 
can also cause cyclic period change. Hence one must take caution when interpreting
the O$-$C diagrams in terms of cyclic changes. Observations so far seem to agree
with the predictions of the orbital period--magnetic modulation period relation (Lanza \&
Rodon\`o 1999) and some of the studied systems were also analysed from this
point of view.

Here we report on multicolour and radial velocity measurements of five close
eclipsing binaries, of which three have never been studied before. In addition, the
sample includes the well-studied quadruple system XY~Leonis, which is one of the
most fascinating cases of  light-travel time effect in a periodic variable star
(Gehlich, Prolls \& Wehmeyer 1972; Yakut et al. 2003), and the triple system VZ~Lib. Besides
revealing new information on these two objects, the observations also allowed a
comparison of our results with previous studies.

\section{Observations and analysis}

The observations were carried out in three observing runs at the Siding Spring
Observatory, Australia. We took two-colour photometry and optical spectroscopy on
six consecutive nights between June 28 and July 4, 2004 for V870~Ara, QY~Hya, VZ~Lib and
DX~Tuc. Data for XY~Leo were gathered on 4 nights in February, 2004 and 7 nights in
March, 2005. The photometry was done using the 40-inch telescope of the Australian
National University (ANU) in Siding Spring, equipped with the Imager CCD detector
(2148$\times$2048 pixels) and Johnson $B$ and $V$ filters. The exposure times ranged
from 3 to 15 seconds, depending on the target brightness and weather conditions.  For
the optical spectroscopy, we used the ANU 2.3-m telescope and the Double Beam
Spectrograph, recording the second order spectra of the 1200 mm$^{-1}$ grating, which
gave a nominal spectral resolution of $\lambda/\Delta \lambda\approx7000$ at the
H$\alpha$ line. We used only the red beam, covering $\sim1000$ \AA\ between the sodium D
doublet and the H$\alpha$ line.

\begin{table}
\begin{center}
\caption{\label{mins} New times of primary (I) and secondary (II) minima for the target stars.}
\begin{tabular}{|lll|}
\hline
$T_{\rm min}$ [HJD] & $T_{\rm min}$ [HJD] & $T_{\rm min}$ [HJD]\\
\hline
 & XY~Leo  & \\
2453412.1839 II & 2453415.0236 II & 2453416.1621 II\\
2453413.0367 II & 2453415.1671 I & 2453435.0538 I\\
2453413.1793 I & 2453416.0209 I & 2453436.0479 II\\
\hline
& VZ~Lib  & \\
2453189.0102 II & 2453190.9776 I & \\
\hline
& DX~Tuc  & \\
2453187.1669 II & 2453188.2989 II & 2453189.2421 I\\
\hline
& QY~Hya  & \\
2453186.9081 I & 2453187.9276 II & \\
\hline
& V870~Ara  & \\
2453185.1379 I & 2453195.1314 I & 2453196.1325 II\\
\hline
\end{tabular}
\end{center}
\end{table}

All data were reduced in a standard fashion using IRAF\footnote{IRAF is distributed by
the National Optical Astronomy Observatories, which are operated by the Association of
Universities for Research in Astronomy, Inc., under cooperative agreement with the
National Science Foundation.}. Photometric reductions included bias and flat-field
correction, while instrumental magnitudes were determined with aperture photometry,
because the stellar fields  around the programme stars were quite empty, so that
accurate psf-fitting was not possible. The instrumental data were tied to the standard
system by observing equatorial photometric standard stars from the list of Landolt
(1992), except for XY~Leo, which was only observed under non-photometric conditions.
The estimated uncertainty of the absolute values of the $B$ and $V$ magnitudes is about
$\pm$0.01 mag.

From the light curves we determined epochs of minimum light by fitting low-order
polynomials. We measured 19 new times of minimum, which are collected in Table\
\ref{mins}. The typical uncertainty is about $\pm3\times 10^{-4}$ d. With these we updated the
O--C diagrams that were based on all published observations in the literature. For
V870~Ara, QY~Hya and DX~Tuc we could derive a more accurate orbital period.

The spectra were also reduced with standard IRAF tasks, including bias and
flat-field corrections, cosmic ray removal, extraction of one-dimensional spectra,
wavelength calibration and continuum normalization. The exposure times ranged from 2
to 5 minutes, and NeAr spectral lamp exposures were regularly taken to monitor the
wavelength shifts of the CCD spectra.  Radial velocities were determined by
cross-correlation, with IAU radial velocity standards  HD~187691 (F8V, $V_{\rm
r}=-0.2$ km~s$^{-1}$) and HD~80170 (K2III, $V_{\rm r}=0.5$ km~s$^{-1}$).  The
calculated cross-correlation functions (CCFs) were fitted with two- or
three-component Gaussians (in case of detected tertiary components), whose centroids
gave the radial velocities for each components. In most cases the Gaussian
approximation gave reasonably good fits for the central parts of the CCFs, while
departures from the Gaussian shape typically occured only 200--300 km~s$^{-1}$  away
from the maxima. 

\begin{figure}
\begin{center}
\leavevmode
\includegraphics[width=8.5cm]{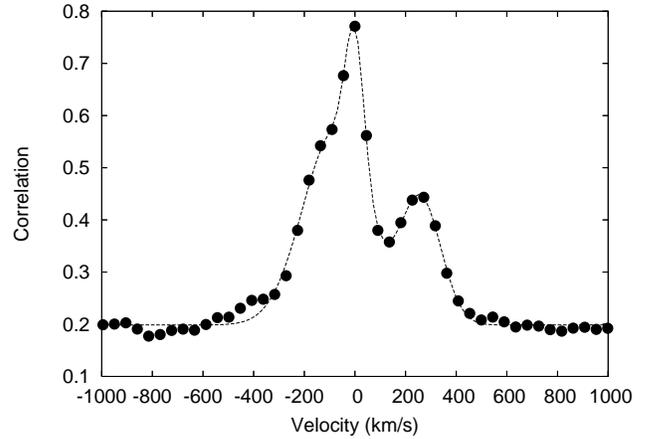}
\end{center}
\caption{A simulated CCF profile and the three-component Gaussian fit (see text for
details).}
\label{ccftest}
\end{figure}

The shape of the CCF clearly showed the presence of a tertiary component in three of
the target stars. In those cases we recorded its mean radial velocity, which can be
used in the future to detect orbital motion around the barycentre of the triple
system. The binary radial velocity curves were fitted with sine curves, which gave
the mean velocity of the binary ($V_\gamma$), the orbital velocities and the
mass-ratio of the components ($K_1$, $K_2$, $q$). These were then used as fixed
input parameters for the light curve models.

The rms scatter of the derived velocity data around sine-wave fits suggests that the 
typical velocity precision is about $\pm$10 km~s$^{-1}$ per point. Although Rucinski
(2002b) demonstrated that the use of the broadening function (BF) for determination of
radial velocities of contact binaries is superior to the CCF method, we chose to apply
the CCF for two reasons. Firstly, IDL was not available 
to us  and thus we could not use the BF 
codes\footnote{http://www.astro.utoronto.ca/$\sim$rucinski/SVDcookbook.html}. Secondly,
the data have relatively poor spectral resolution and thus the 
CCF approach is sufficient. Nevertheless, besides
estimating probable errors from the data, we have also tested the accuracy of the
velocity measurement method in the following way. First, a theoretical spectrum of a 
$T_{\rm eff}=5900$ K, $\log g = 4.0$ star was calculated with Robert Kurucz's  ATLAS9
code between 5800 \AA~ and 6800 \AA, and resampled to a resolution of $\sim 8000$. Then,
a contact binary broadening profile at quadrature phase ($\phi = 0.25$) was generated by
Slavek Rucinski's WUMA4 code, applying the physical parameters of VZ~Lib, one of our
program stars (see Sect.\ 3.2).  The model spectrum was then convolved with the
broadening profile to get a contact binary model spectrum. A scaled original model
spectrum (scaled to 20\% of the initial flux level) at zero velocity was added to the
contact binary model spectrum to mimic the presence  of the tertiary component. This
binary+tertiary spectrum was then  cross-correlated with the original model spectrum. The
resulting CCF, which is very similar to most of the real CCFs of our program stars, is
shown in  Fig.\ \ref{ccftest}. 

The radial velocities were determined via fitting three-component Gaussians, as
described above. The centroids recovered for the model spectrum are  $a_1 = -80.5$
km~s$^{-1}$, $a_2 = 259.1$ km~s$^{-1}$ and  $a_3 = 0.0$ km~s$^{-1}$, while the model
velocities were initially $v_1 = -84$ km~s$^{-1}$, $v_2 = 252$ km~s$^{-1}$ and $v_3 =
0.0$ km~s$^{-1}$. It is seen that the error of the velocity determination is within
$10$ km~s$^{-1}$ for the binary components and less than that (probably a few
km~s$^{-1}$) for the sharp-lined third component. These are clearly larger than the
$1 - 2$ km~s$^{-1}$ error determined by Rucinski (2002b) for the BF-method, but our
spectra are of lower resolution than those applied by Rucinski (2002b). The numbers
also show that the recovered velocities are shifted to the input values in the same
direction, which means the effect is systematic, affecting both velocity amplitudes. 
We note that Rucinski and co-workers found that CCF tends the reduce the measured
radial velocity amplitudes relative to the BF results (e.g. Rucinski 2004).
Conservatively, we adopt a $\pm 8$ km~s$^{-1}$ error in the derived 
amplitudes $K_1$ and $K_2$ and $\pm 2$ km~s$^{-1}$ per point as the
measurement uncertainty of the radial velocity of the
tertiary component. These errors were added in quadrature to the formal errors 
of the calculated fits.

The light curves were modelled using the 2003 version\footnote{
ftp://ftp.astro.ufl.edu/pub/wilson} of the  Wilson-Devinney (WD) code (Wilson \&
Devinney 1971, Wilson 1994, Wilson \& van Hamme 2003). Besides the spectroscopic
parameters, we determined the effective temperature, $T_{1}$, of the component
eclipsed in zero phase from the $(B-V)-T_{\rm eff}$ calibration of Gray (1992). The
lower limit of the semimajor axis, $A$, was calculated from  $K_1$, $K_2$ and $P$.
These input parameters were kept fixed while running the WD code on the original
individual data points. Limb-darkening coefficients were taken from Diaz-Cordoves,
Claret \& Gimenez (1995), while the bolometric albedos and gravity-darkening
coefficients were set to 0.5 (Rucinski 1969) and 0.32 (Lucy 1967), respectively. The
third light was included in the fitting procedure, except for DX~Tuc and V870~Ara, 
for which we do not find evidence for a third component. In three cases, the
brightness difference of the consecutive maxima (the O'Connell-effect, hereafter
denoted by $\Delta V$ in the $V$ band) suggested the presence of spots, which were 
added to the solution. The astrographic coordinates of the center of a given spot
were fixed after several trials. 

In case of the multiple systems, we also estimated the physical parameters 
of the third bodies using the method described in Borkovits et al. (2002)
and Csizmadia (2005). The light curve solutions yield the third
light in different colours and the colour index of the
tertiary component follows from the colour index
of the primary star as

\begin{equation} 
(B - V)_3 = (B - V)_1 - 2.5\left[ \log(\frac{L_{3,B}}{L_{3,V}}) - \log(\frac{L_{1,B}}{L_{1,V}})\right]
\end{equation}

\noindent where $L_3 \approx 4\pi l_3$ ($l_3$ is the third light) and the luminosities
are given in arbitrary units calculated by the WD-code. The colour index of the 
primary was estimated from the determined effective temperature using
the temperature--$(B-V)$ relations by Flower (1996).
With only two colours we could not identify the location of the
tertiary star in a two-colour diagram. Therefore, we assumed that the
tertiary component is a main-sequence star. Then its spectral type can be
estimated from the colour index using the Bessell (1990) tables, yielding
an approximate mass estimate.

All the standardized light curves and radial velocity data are available electronically
at the CDS via anonymous ftp to {\tt cdsarc.u-strasbg.fr (130.79.128.5)} or via {\tt
http://cdsweb.u-strasbg.fr/cgi-bin/qcat?J/\\AA/.../...}. 

\section{Results}

\subsection{XY~Leonis}

\begin{table*}
\caption{Physical parameters of XY~Leo. Parameters from the literature were taken
from Barden (1987 -- spectroscopy) and Yakut et al. (2003 -- light curve fit and
absolute parameters). Values in italic denote fixed input parameters. HJD$_3$
and $V_3$ refer to mean epoch and radial velocity of the third component.}
\begin{footnotesize}
\begin{center}
\begin{tabular}{|l|l|l||l|l|l|}
\hline
Parameter & This paper & Literature & Parameter & This paper & Literature\\
\hline
\hline
Spectroscopy &             &           &             &          &    \\
$A\sin i$ [R$_{\odot}$] & 1.80$\pm$0.11 & 1.84$\pm$0.03 & $K_{1}$ [km/s] & 203.7$\pm$9.1 & 204.7$\pm$2.5\\
$V_{\gamma}$ [km/s] & $-$36.7$\pm$5.1 & $-$51.8$\pm$2.2 & $K_{2}$ [km/s] & 122.0$\pm$10.0 & 124.1$\pm$2.8\\
HJD$_{3}$ & 2453456.5 & -- & $V_3$ [km/s] & $-$70.0$\pm$7.5 &\\
\hline
Light curve & fit                      &          &             &          &    \\
$i$ $[^{\circ}$] & 67.1$\pm$0.1 & 68$\pm$1 & $\Omega_{1}$ & 4.7206$\pm$0.0037 & 4.71$\pm$0.01\\
$f$ & 5.2\% & 6.7\% & $\Omega_{2}$& 4.7206 & 4.71\\
phase shift & 0.0027$\pm$0.0002 & -- & $(\frac{L_{1}}{L_{1}+L_{2}})_{B}$ & 0.553$\pm$0.009 & 0.507$\pm$0.043\\
$q$ & {\it 1.66($\pm$0.23)} &{\it 1.64} & $(\frac{L_{1}}{L_{1}+L_{2}})_{V}$ & 0.522$\pm$0.007 & 0.483$\pm$0.043\\
$T_{1}$ [K] &{\it 4800} &{\it 4850} & $(l_{3})_{B}$ & 0.051$\pm$0.005 & 0.019$\pm$0.007\\
$T_{2}$ [K] & 4361$\pm$8 & 4524$\pm$14 & $(l_{3})_{V}$ & 0.068$\pm$0.004 & 0.059$\pm$0.007\\
$(B-V)_1$ & 1.02 &  -- & $(B-V)_3$ & 2.02 & -- \\
\hline
Absolute & parameters               &           &            & &               \\
$M_{1}$ [M$_{\odot}$] & 0.46$\pm$0.06 & 0.50$\pm$0.02 & $M_{2}$ [M$_{\odot}$] & 0.76$\pm$0.15 & 0.82$\pm$0.02 \\
$R_{1}$ [R$_{\odot}$] & 0.66$\pm$0.10 & 0.68$\pm$0.02 & $R_{2}$ [R$_{\odot}$] & 0.83$\pm$0.13 & 0.85$\pm$0.02\\
$L_1$  [L$_\odot$] & 0.21$\pm$0.07 & 0.226&  $L_2$ [L$_\odot$] & 0.22$\pm$0.08& 0.267\\
$(M_{\rm bol})_1$ & 6.43$\pm$0.23& 6.4 &  $(M_{\rm bol})_2$ & 6.34$\pm$0.23 & 6.2\\
$(M_{\rm V})_1$ & 6.89$\pm$0.23 & 6.7& $(M_{\rm V})_2$ & 6.96$\pm$0.23& 6.7\\
Sp. type (3) & M & & & & \\
$d$ [pc] & 51$\pm$6 & 63$^{+8}_{-6}$ & & & \\ 
\hline
\end{tabular}
\end{center}
\label{leopars}
\end{footnotesize}
\end{table*}

\begin{figure*}
\begin{center}
\leavevmode
\includegraphics[width=8cm]{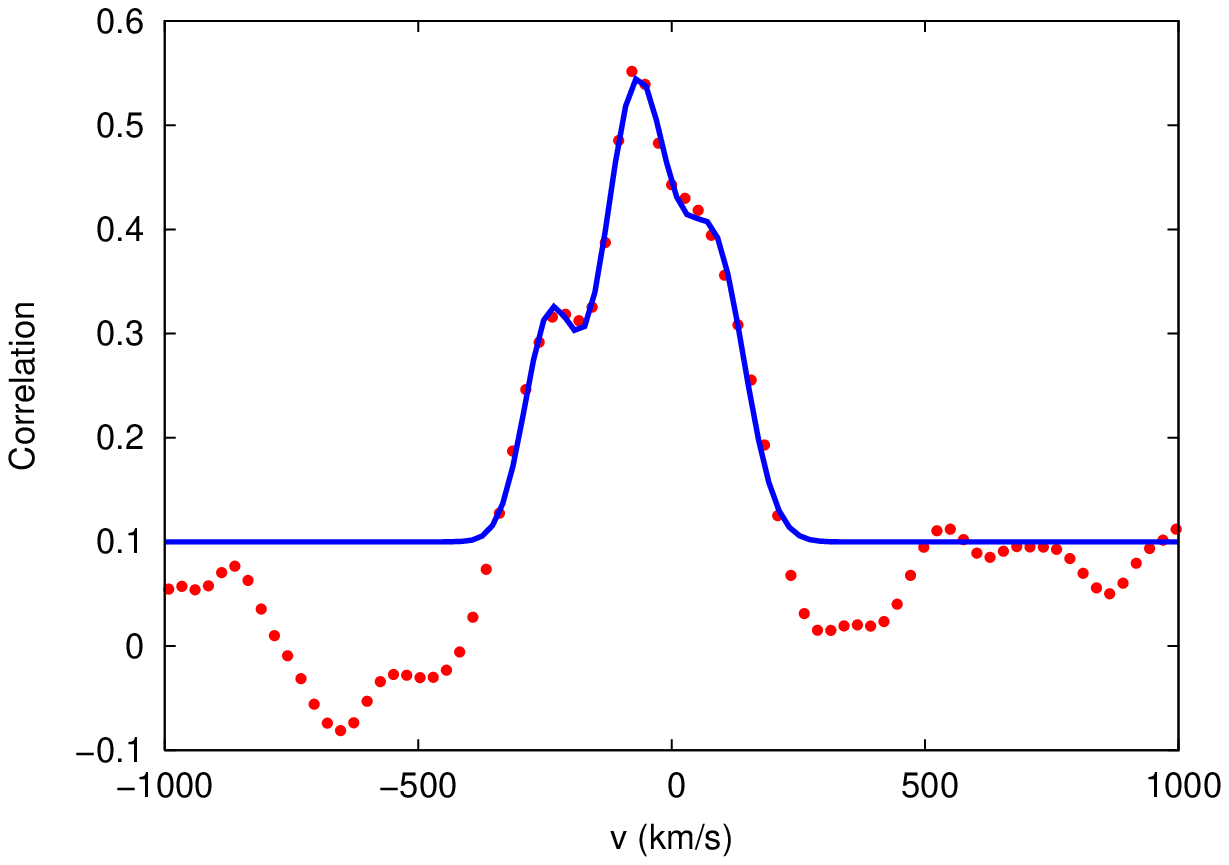}
\includegraphics[width=8cm]{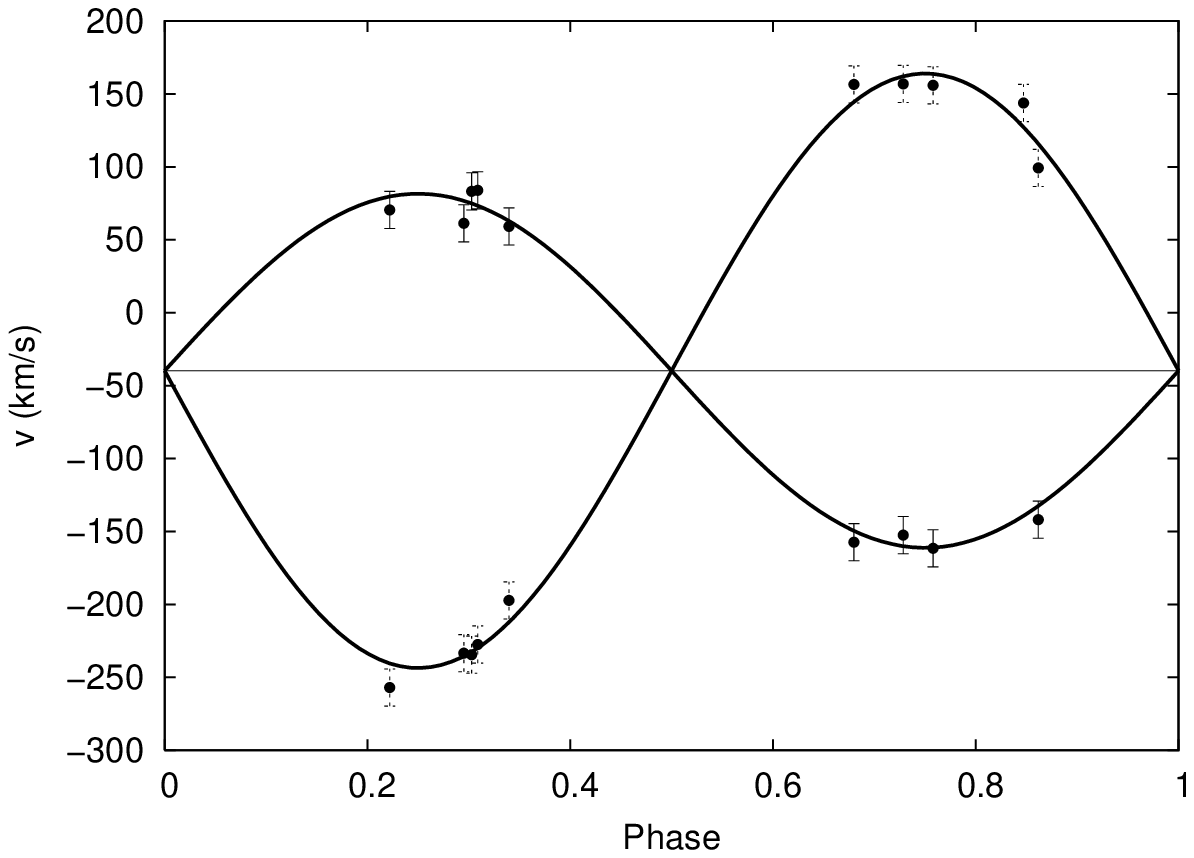}
\end{center}
\caption{{\it Left panel:} the CCF-profile of XY~Leo in phase 0.75 with the three fitted Gaussians.
{\it Right panel:} radial velocity curves of the eclipsing components of XY~Leo. } 
\label{leoccf}
\end{figure*}

XY~Leo is a well-studied contact binary which has long been known for its cyclic
period variations (Gehlich, Prolls \& Wehmeyer 1972; Yakut et al. 2003). In stark
contrast to the multitude of photometric studies (Yakut et al. 2003 and references
therein), only two spectroscopic studies were aimed at measuring radial velocity
curves (Hrivnak et al. 1984; Barden 1987). XY~Leo has four components. One pair
is a W-type contact binary (i.e. the smaller star is the hotter one) with very short
orbital period ($P\approx0.28$ d) and late spectral type (Kn).  It is a
chromospherically active variable (Vilhu \& Rucinski 1985), but the activity  is
largely dominated by the BY~Dra-type binary component that consists of two  red
dwarf stars (Barden 1987).

Spectra taken in early 2004 showed a similar, though slightly weaker and unresolved
H$\alpha$  emission than the one depicted in fig. 1 of Barden (1987). The
cross-correlation  profiles of the spectra clearly indicated the presence of a third
component (left panel in Fig.\ \ref{leoccf}), so we fitted three Gaussians to
measure radial velocities of the eclipsing pair (right panel in Fig.\ \ref{leoccf}).
The determined spectroscopic elements (Table\ \ref{leopars}) agree within the
errorbars with the corresponding parameters of Barden (1987), except the
$\gamma$-velocity. Since our observations were taken 19 years later than those of
Barden (1987), which is close to the orbital period of the wide binary, this shift
in $V_\gamma$ might be due to the strong third light that distorts the whole CCF
profile. Nevertheless, the agreement in the measured amplitudes suggests that the 
resulting velocity shift caused by the distorted CCF is very similar for both
components.   

\begin{figure}
\begin{center}
\leavevmode
\includegraphics[width=8cm]{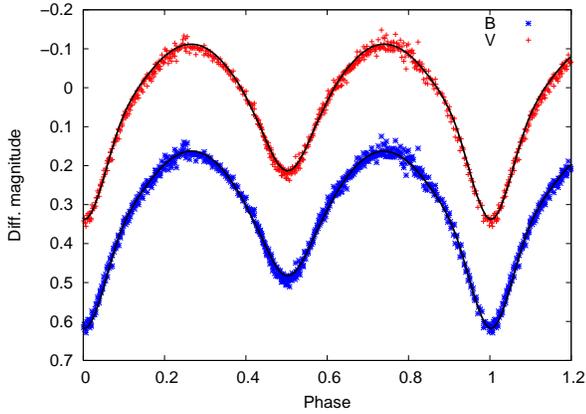}
\end{center}
\caption{Observed and fitted light curves of XY~Leo.}
\label{leofit}
\end{figure}

The light curves (Fig.\ \ref{leofit}) do not show O'Connell-effect, thus  we
did not include spots in the light curve fit. In this case we had only
differential light curves, so the fixed temperature of the primary component was
estimated from the spectral type. The results (Table\ \ref{leopars}) are quite
similar  to those of Yakut et al. (2003). The derived colour index of the third component
is very red at $(B-V)_3=2.02$ mag, being consistent with a BY~Dra-like M-dwarf
binary (Barden 1987). For calculating the distance we used the maximum
$V$-brightness of the stars listed in ESA (1997).

\begin{figure}
\begin{center}
\leavevmode
\includegraphics[width=8cm]{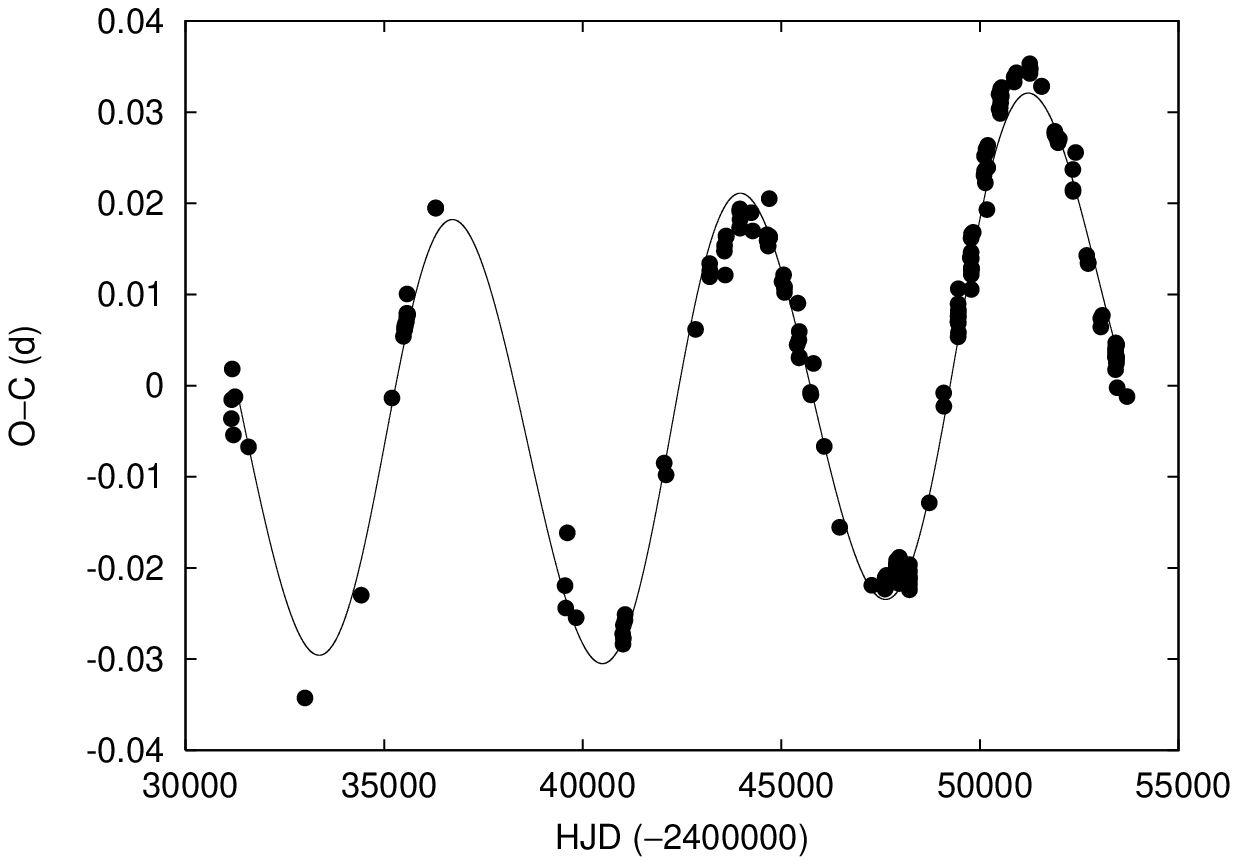}
\includegraphics[width=8cm]{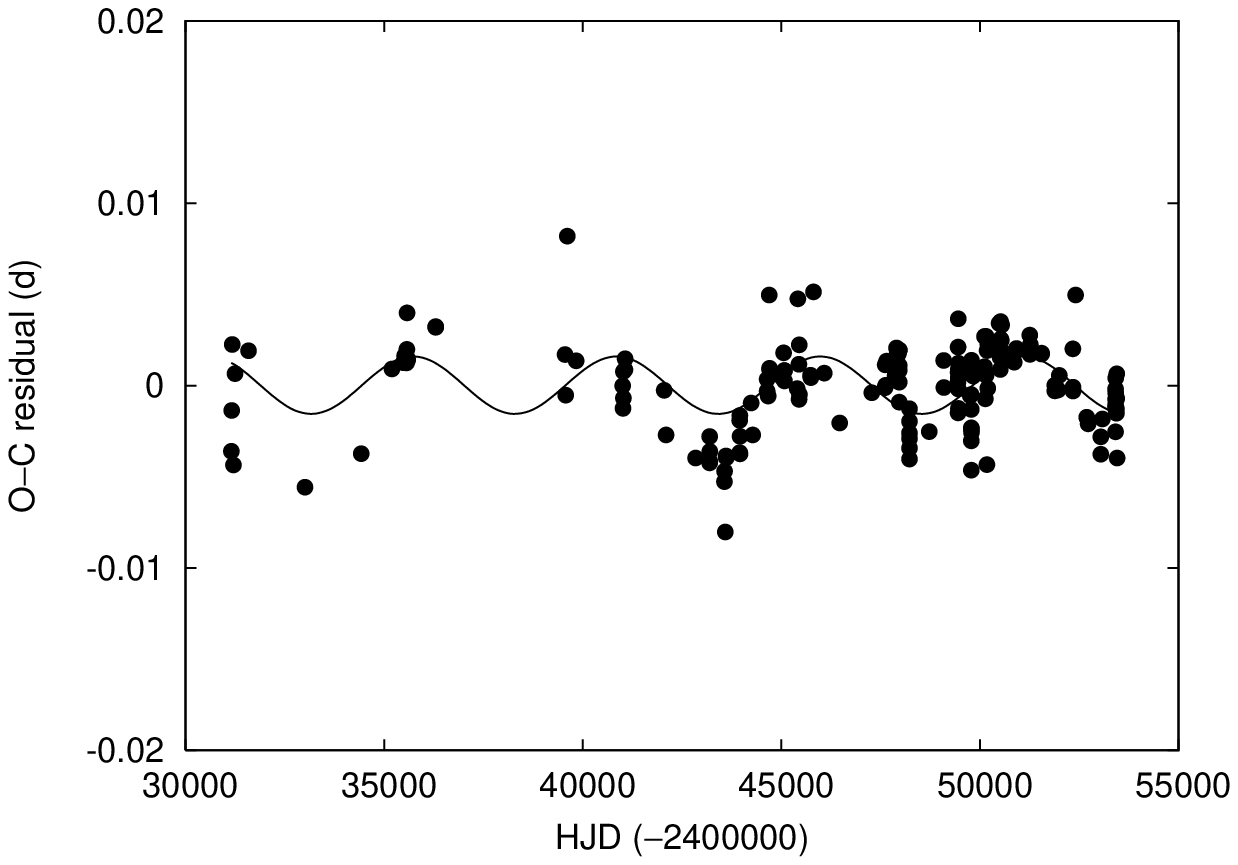}
\end{center}
\caption{{\it Top panel:} the O$-$C diagram of XY~Leo with the theoretical fit.
{\it Bottom panel:} the residuals and the $\sim$5100 d secondary periodicity.}
\label{leooc}
\end{figure}

With the most extended coverage of the period change, we also updated the O$-$C
diagram. Since the publication of the Yakut et al. (2003) analysis, new moments of
minima have been published by Agerer \& H\"ubscher (2003), Drozdz \& Og\l oza
(2005), H\"ubscher (2005), H\"ubscher, Paschke \& Walter (2005) and Nelson (2006).
We also added a few epochs from Tsessevitch (1954a), which were not included in the
analysis of Yakut et al. (2003) but were important in stretching the time coverage
as far back as possible. The full O$-$C diagram of XY~Leo, plotted in the
top panel of  Fig.\ \ref{leooc}, was calculated using the ephemeris: HJD$_{\rm
min}=2435484.0283+0.28410260\times E$. Then we simultaneously fitted a parabolic and
a light-time term to calculate the rate of the linear period change  and the orbital
elements of the perturbing binary. For this, we used the same code as Ribas, Arenou
\& Guinan (2002), kindly provided by Dr. I. Ribas.

The results are summarized in Table\ \ref{leolite}. In general, the parabolic term
represents a continuous period variation that might be explained by some
mass-transfer between the components. The period changing rate of XY~Leo
$\frac{1}{P} \frac{dP}{dt} = 2.85 \times 10^{-8}$ yr$^{-1}$, which can be translated
to a mass-transfer rate of $-9.6 \times 10^{-9}$ M$_\odot$/yr. This period
variation  rate is in the usual range of W~UMa systems (see Qian 2001ab; Borkovits
et al. 2005 -- XY~Leo was not included in these studies). The secular orbital period
increase seems to support what  Qian (2001a) suggested, namely that W-type systems
with $q<0.4$ and $q>0.4$ tend to show secular decrease and increase,  respectively. 

The light-time parameters in Table\ \ref{leolite} give very similar orbital
elements for the quadruple system to all the previous investigations  (Gehlich,
Prolls \& Wehmeyer 1972; Pan \& Cao 1998; Yakut et al. 2003), so that we do not
re-iterate the discussion of the hierarchic component. However, the residuals of the
O$-$C fit deserve further discussion, because the 201 s rms is larger than the
typical observational uncertainties. An analysis of the long-term period change of
the triple W~UMa VW~Cep revealed secondary cycles in the O$-$C diagram that were
interpreted as possible evidence for magnetic activity (Kasz\'as et al. 1998). More
recently, Borkovits et al. (2005) investigated complex period variations of five
W~UMa-type binaries, finding some common features. These include secular period
variation at a constant rate and a low-amplitude modulation with periods around
18--20 yr in four of the five cases. These stars all have spectral type later than
F8 (so they have convective envelopes), while the fifth one is an A-type contact
binary. XY~Leo is a K-type star, therefore the secondary periodicity of the O$-$C
diagram is in agreement with the picture suggested by Borkovits et al. (2005) that
all contact binaries later than F8 have low-amplitude cyclic period changes. These
can be interpreted as indirect evidence of magnetic cycles in late-type overcontact
binaries that are analogous to the observed activity cycles in RS~CVn systems (Hall
1989). The theoretical framework involves the interchange of magnetic and kinetic
energy  (Applegate 1992; Lanza \& Rodon\`o 2002), predicting time-scales from
several years to decades. 

\begin{table}
\caption{Parameters of the O$-$C fit. The standard errors in the last digits
are shown in parentheses.}
\begin{center}
\begin{tabular}{ll}
\hline
Parameter & value\\
\hline
$A_{\rm O-C}$ [d] & 0.02430(1)\\ 
$e$ & 0.11(1)\\
$\omega$ [$^\circ$] & 18.5(5)\\
$P_3$ [yr] & 19.651(2) \\
$T_{\rm periastron}$ [HJD] & 2446119(11)\\
$P_1$ [d] & 0.28410295(1)\\
$dP/dt$ [s/d] & 1.92(6)$\times$10$^{-6}$\\
\hline
\end{tabular}
\end{center}
\label{leolite}
\end{table}

The bottom panel of Fig.\ \ref{leooc} shows the best-fit sine wave to the
residuals of the O$-$C. Introducing this extra component decreases the rms to
$\sim$175 s, which represents a slight but detectable improvement and is in
agreement with the typical photometric noise from starspots (Kalimeris,
Rovithis-Livaniou \& Rovithis 2002). The Fourier spectrum of the residuals shows
that the highest peak has a S/N ratio  (Breger et al. 1993) of 3, thus its
significance is marginal.  Nevertheless, the corresponding modulation period
($\sim$5100 d or 14 yr) and  amplitude (0.0016 d) are very similar to those 
found by Borkovits et al. (2005) for AB~And, OO~Aql, V566~Oph and U~Peg.  We
therefore conclude that XY~Leo is likely to bear signatures of  short-period
magnetic cycles in W~UMa-type variables, but several years of further continuous
eclipse timings are needed to improve the significance of the detection.

\subsection{VZ~Librae}

\begin{table*}
\caption{Physical parameters of VZ~Lib. Parameters from the literature were taken
from Lu, Rucinski \& Og\l oza (2001 -- spectroscopy) and Zola et al. 
(2004 -- light curve fit and absolute parameters). Values in italic denote 
fixed input parameters. HJD$_3$
and $V_3$ refer to mean epoch and radial velocity of the third component.}
\begin{center}
\begin{tabular}{|l|l|l||l|l|l|}
\hline
Parameter & this paper & Literature & Parameter & this paper &  Literature\\
\hline
\hline
Spectroscopy &             &           &             &              & \\
$A\sin i$ [R$_{\odot}$] & 2.38$\pm$0.12 & 2.53$\pm$0.06 & $K_{1}$ [km/s] & 84.1$\pm$8.2 & 68.5$\pm$3.8\\
$V_{\gamma}$ [km/s] & $-$51.1$\pm$1.9 & $-$31.1$\pm$2.3 & $K_{2}$ [km/s] & 252.2$\pm$8.3 & 289.2$\pm$4.6\\
HJD$_3$ & 2453189.4 & -- & $V_3$ [km/s] & $-$4.8$\pm$3 & \\
\hline
Light curve & fit                  &          &             &           &   \\
$i$ [$^{\circ}$] & 88.4$\pm$1.0 & 80.3$\pm$0.5 & $r_1^{\rm side}$ & 0.4908$\pm$0.0009 & 0.5122$\pm$0.0012\\
phase shift & $-$0.0004$\pm$0.0003 & $-$0.0030$\pm$0.0004 & $r_2^{\rm side}$ & 0.2905$\pm$0.0008 & 0.2655$\pm$0.0011\\
$q$ &{\it 0.33$(\pm$0.04)} &{\it 0.255} & $r_1^{\rm back}$ & 0.5195$\pm$0.0011 & 0.5378$\pm$0.0015\\
$T_{1}$ [K] &{\it 5770} &{\it 5920} & $r_2^{\rm back}$ & 0.3303$\pm$0.0014 & 0.3022$\pm$0.0020\\
$T_{2}$ [K] & 5980$\pm$12 & 6030$\pm$21 & $f$ & 19.4\% & 13\%\\
$\Omega_{1}$ & 2.498$\pm$0.003 & 2.344$\pm$0.004 & $(\frac{L_{1}}{L_{1}+L_{2}})_{B}$ & 0.686$\pm$0.002 & 0.757$\pm$0.001\\
$\Omega_{2}$ & 2.498 & 2.344 & $(\frac{L_{1}}{L_{1}+L_{2}})_{V}$ & 0.545$\pm$0.009 & 0.759$\pm$0.002\\
$r_1^{\rm pole}$ & 0.4558$\pm$0.0007 & 0.4732$\pm$0.0009 & $l_{3}$ (B) & 0.192$\pm$0.007 & 0.011$\pm$0.005\\
$r_2^{\rm pole}$ & 0.2777$\pm$0.0007 & 0.2545$\pm$0.0010 & $l_{3}$ (V) & 0.209$\pm$0.004 & 0.043$\pm$0.009\\
$(B-V)_1$ & 0.63 & -- & $(B-V)_3$ & 0.73 & -- \\
\hline
Absolute & parameters      &         &           &            &               \\
$M_{1}$ [M$_{\odot}$] & 1.06$\pm$0.06 & 1.480$\pm$0.068 & $M_{2}$ [M$_{\odot}$] & 0.35$\pm$0.03 & 0.378$\pm$0.034 \\
$R_{1}$ [R$_{\odot}$] & 1.17$\pm$0.05 & 1.34$\pm$0.02& $R_{2}$ [R$_{\odot}$] & 0.72$\pm$0.03 & 0.69$\pm$0.01\\
$L_1$ [L$_\odot$] & 1.36$\pm$0.19 & --&  $L_2$ [L$_\odot$] & 0.59$\pm$0.05& --\\
$(M_{\rm bol})_1$ & 4.39$\pm$0.11& -- &  $(M_{\rm bol})_2$ & 5.29$\pm$0.11 &--\\
$(M_{\rm V})_1$ & 4.60$\pm$0.11 & --& $(M_{\rm V})_2$ & 5.48$\pm$0.11& --\\
Sp. type (3) & G7 & -- & $M_3$ [M$_\odot$] & 0.90 & --\\
$d$ [pc] & 171$\pm$8 & -- & & & \\
\hline
\end{tabular}
\end{center}
\label{libpars}
\end{table*}

\begin{figure*}
\begin{center}
\leavevmode
\includegraphics[width=8cm]{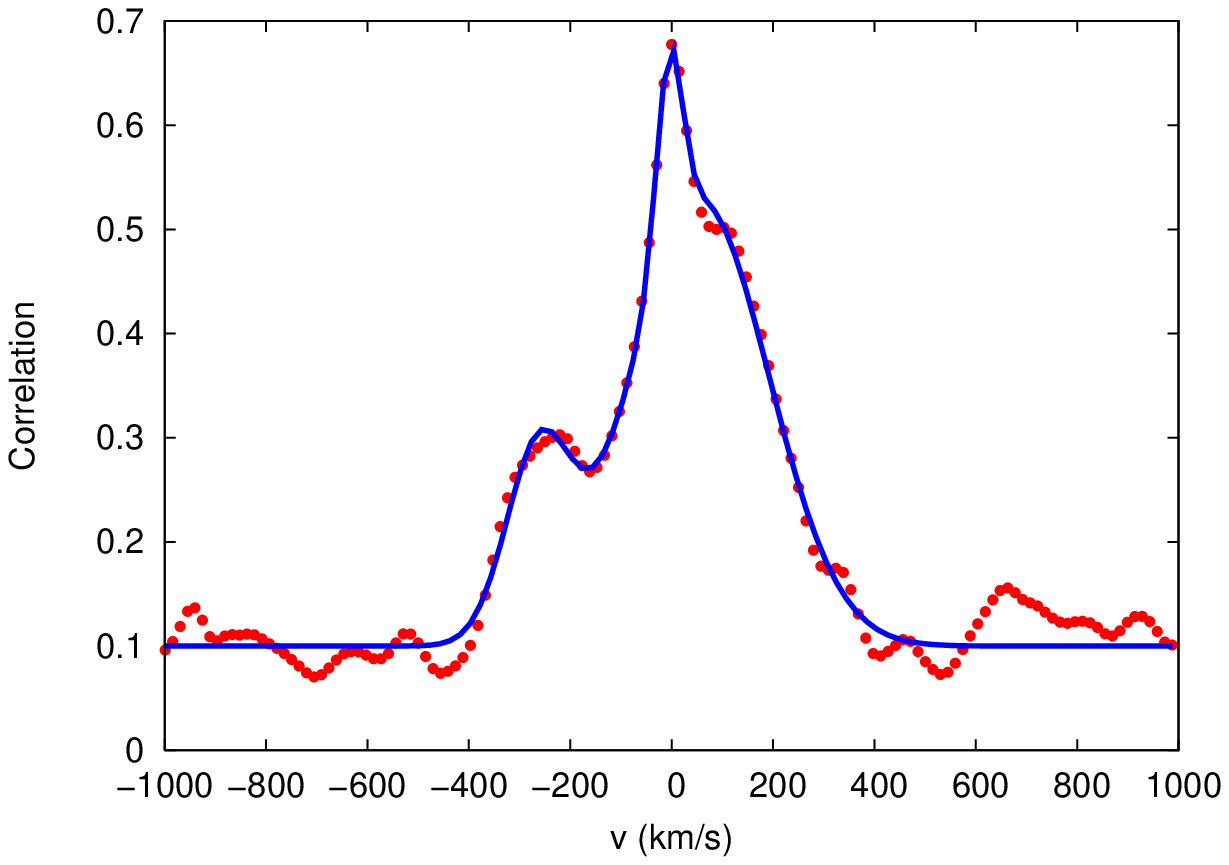}
\includegraphics[width=8cm]{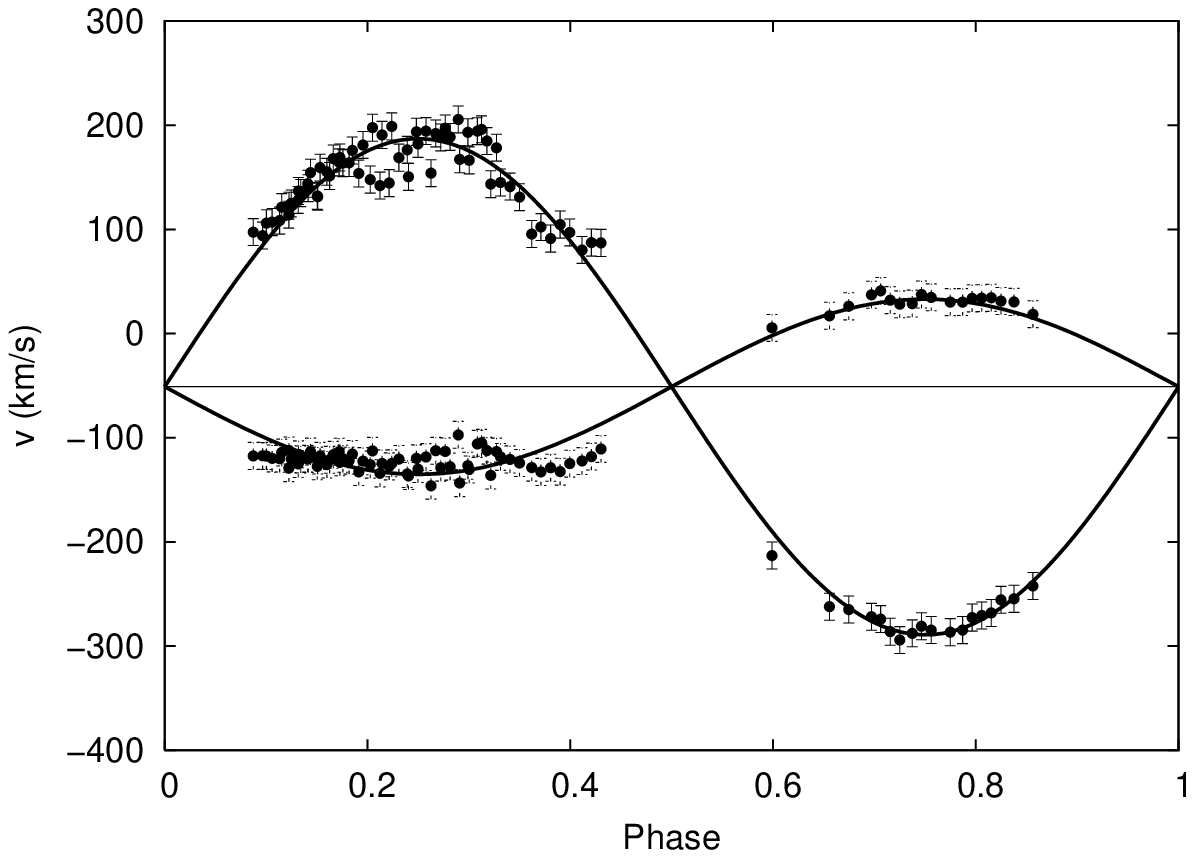}
\end{center}
\caption{{\it Left panel:} the CCF-profile of VZ~Lib in phase 0.75 with the 
three fitted Gaussians. {\it Right panel:} radial velocity curves of the 
eclipsing components of VZ~Lib.}
\label{libccf}
\end{figure*}

The variability of the F5-type star VZ~Lib was discovered by Hoffmeister (1933), who
gave neither  period nor classification but recognized its low-amplitude eclipsing
nature. Tsessevich (1954b) classified the star as  a W~UMa-type variable with a
period slightly over 8 hours. Claria \& Lapasset (1981) reported the first
photoelectric light curve which showed a difference in the eclipse depths for the
primary and the secondary minima of about 0.1 mag. Interestingly, more recent
observations, including ours and those of Zola et al. (2004) find more similar
minima with differences only 0.02--0.03 mag (Zola et al. 2004) and $<$0.01 mag (this
paper) in $V$. There is a relatively bright tertiary component in the system
discovered by Lu, Rucinski \& Og\l oza (2001). Besides finding radial velocity
variations of the third star up to 40 km~s$^{-1}$ over a period of 1200 days, they
derived a luminosity ratio of $L_3/L_{12}=0.20\pm0.04$. Contrary to this, Zola et
al. (2004) arrived to only  a few (4-5) percent of third light contribution, which
was left unexplained. 

The cross-correlation profiles of our spectra clearly indicated the presence of a
third, narrow-lined companion (left panel in Fig.\ \ref{libccf}). Its mean radial
velocity is $-4.8\pm3$ km~s$^{-1}$, which is larger than any of the velocities in
fig. 5 of Lu, Rucinki \& Og\l oza (2001), ranging from $-9$ to $-50$ km~s$^{-1}$.
Apart from pushing up the full velocity range of the third component well over $40$
km~s$^{-1}$, this value strongly suggests that the 1200 d of observations by Lu,
Rucinski \& Og\l oza (2001) was shorter than the orbital period of the third
component. 

\begin{figure}
\begin{center}
\leavevmode
\includegraphics[width=8cm]{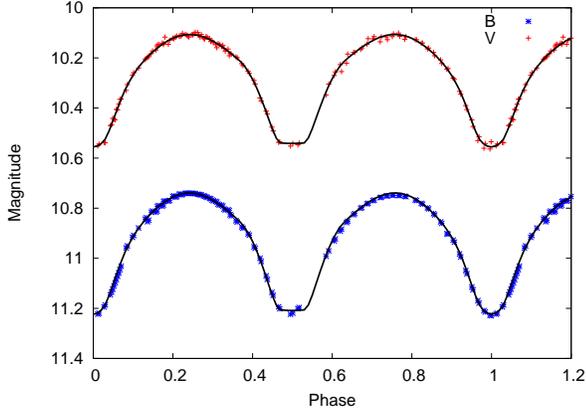}
\end{center}
\caption{Observed and fitted light curves of VZ~Lib.}
\label{libfit}
\end{figure}

The radial velocity curve (right panel in Fig.\ \ref{libccf}) yields a higher
mass-ratio than that of Lu, Rucinski  \& Og\l oza (2001), but it is still compatible
with the earlier result of $q=0.24\pm0.07$ that was based on a relatively poor radial
velocity solution. The light curves (Fig.\ \ref{libfit}) are very similar to those of
Zola et al. (2004), except that the O'Connell-effect was not detectable in 2004.
Consequently, our light curve fit did not include spots on any of the components.

A comparison of the determined parameters with the published ones shows a relatively
good agreement (Table\ \ref{libpars}). The differences can largely be traced back to
our larger mass-ratio and the different amount of the third light. The strong
tertiary peak in the CCF-profile supports the $\sim$20\% luminosity contribution of
the third component, which is also close to the value determined by Lu, Rucinski \&
Og\l oza (2001). Since the Zola et al.  light curve model is not consistent with
this, we are confident that our light curve solution is more compatible with the
available spectroscopic information. To remove most of the ambiguities, a better
defined radial velocity  curve is highly desirable, preferentially from
higher-resolution and S/N spectra.

Because the tertiary component is comparable in luminosity (and perhaps in mass) to
the eclipsing pair, one can expect measurable light-time effect in the period change.
Since the study of Claria \& Lapasset (1981) only the Hipparcos photometry, Zola et
al. (2004) and Krajci (2006) presented new epochs of minimum light, which is hardly
enough to reconstruct the hypothetic cyclic period changes. On the other hand, the
ASAS-3 project (Pojmanski 2002) has been observing this system since 2001 and
produced  259 $V$-band points scattered between JD 2451918--2453277. We combined the
recent data as follows. 

\begin{figure}
\begin{center}
\leavevmode
\includegraphics[width=8cm]{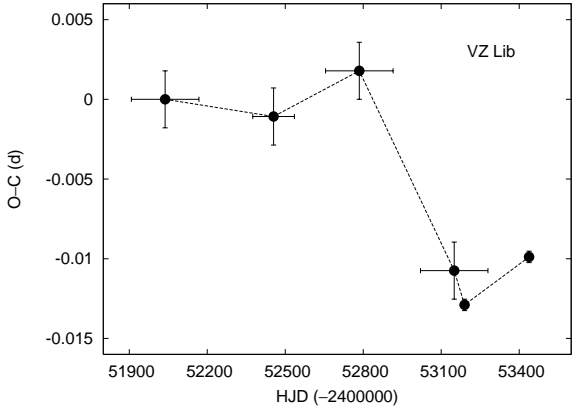}
\end{center}
\caption{The O$-$C diagram of VZ~Lib between 2000--2006 (see text for details).}
\label{liboc}
\end{figure}

First we adopted the linear ephemeris of Zola et al. (2004): $E_0=2452727.4047$,
$P=0.3582580$ d. Then we cut the ASAS data into 4 equal subsets, which correspond
to four observing seasons since the beginning of the project. We phased all the subsets
using the given ephemeris and the resulting phase diagrams yielded the phase-shifts 
of the primary minimum. In Fig.\ \ref{liboc} we plot the results (multiplied by the
period), where the last two points were calculated from our primary minimum and 
that of Krajci (2006). 

The shape of the O$-$C diagram indicates measurable phase changes 
over the last 6 years. This may be interpreted as possible evidence for period 
variations due to light-time effect in the triple system, although other 
mechanisms, like magnetic activity or sudden mass exchange via large flares, 
cannot be ruled out either. Assuming light-time effect, both the long-term 
radial velocity curve (Lu, Rucinski \& Og\l oza 2001) and Fig.\
\ref{liboc} suggest that the orbital period of the third component is longer 
than 1200--1500 d. Regular future eclipse timings will therefore be crucial 
in sorting out different mechanisms of period change and, ultimately, 
improving our understanding of the system.

\subsection{DX~Tucanae}

The variability of DX~Tuc was discovered by the Hipparcos satellite and it was 
classified as an F7-type contact binary (ESA 1997, Kazarovets et al. 1999). Pribulla,
Kreiner \& Tremko (2003) listed the star in their catalog of field contact binary
stars. Selam (2004), based on the Fourier-decomposition of the Hipparcos light curve,
put it among the 64 genuine W~UMa-type variables, while 
Pribulla \& Rucinski (2006) did not find any indication for multiplicity (caused
mainly by the lack of available data in the literature).

\begin{table}
\caption{Physical parameters of DX~Tuc.}
\setlength{\tabcolsep}{2mm}
\begin{center}
\begin{tabular}{|llll|}
\hline
Parameter & Value & Parameter & Value \\
\hline
\hline
Spectroscopy &   &    & \\
$A\sin i$ [R$_{\odot}$] & 2.24$\pm$0.12 &          $K_{1}$ [km/s] & 66.8$\pm$8.1\\
$V_{\gamma}$ [km/s] & 25.4$\pm$0.8 &  $K_{2}$ [km/s] & 233.8$\pm$8.1\\
\hline
Light curve & fit        && \\
$i$ [$^{\circ}$] & 62.3$\pm$0.2        & $r_1^{\rm side}$ & 0.5059$\pm$0.0006\\
phase shift & 0.0022$\pm$0.0002      & $r_2^{\rm side}$ & 0.2850$\pm$0.0006\\
$q$ & {\it 0.29}($\pm$0.04)            & $r_1^{\rm back}$ & 0.5348$\pm$0.0008\\
$T_{1}$ [K] &{\it 6250}                & $r_2^{\rm back}$   & 0.3279$\pm$0.0010\\
$T_{2}$ [K] & 6182$\pm$37            & $f$ & 14.9\%\\
$\Omega_{1}$ & 2.408$\pm$0.002       & $(\frac{L_{1}}{L_{1}+L_{2}})_{B}$ & 0.756$\pm$0.002\\
$\Omega_{2}$ & 2.408                 & $(\frac{L_{1}}{L_{1}+L_{2}})_{V}$ & 0.755$\pm$0.001\\  
$r_1^{\rm pole}$ & 0.4676$\pm$0.0004 & $l_{3}$ (B) &{\it 0.0}\\
$r_2^{\rm pole}$ & 0.2721$\pm$0.0005 & $l_{3}$ (V) &{\it 0.0}\\
\hline
Spot & parameters  & &  \\
Co-lat. [deg] & {\it 92} & Rad. [deg] & {\it 26}\\
Long. [deg] & {\it 175} & $T_{\rm fact}$ & 0.97$\pm$0.01\\
\hline
Absolute &  parameters       && \\
$M_{1}$ [M$_{\odot}$] & 1.00$\pm$0.03 & $M_{2}$ [M$_{\odot}$] & 0.30$\pm$0.01 \\ 
$R_{1}$ [R$_{\odot}$] & 1.20$\pm$0.04& $R_{2}$ [R$_{\odot}$] & 0.71$\pm$0.02\\
$L_1$ [L$_\odot$] & 1.97$\pm$0.25 & $L_2$ [L$_\odot$] & 0.66$\pm$0.04\\
$(M_{\rm bol})_1$ & 3.98$\pm$0.10 &  $(M_{\rm bol})_2$ & 5.17$\pm$0.10\\
$(M_{\rm V})_1$ & 4.13$\pm$0.10 &  $(M_{\rm V})_2$ & 5.34$\pm$0.10\\
$d$ [pc] & 128$\pm$6 & & \\
\hline
\end{tabular}
\end{center}
\label{tucpars}
\end{table}

\begin{figure}
\begin{center}
\leavevmode
\includegraphics[width=8cm]{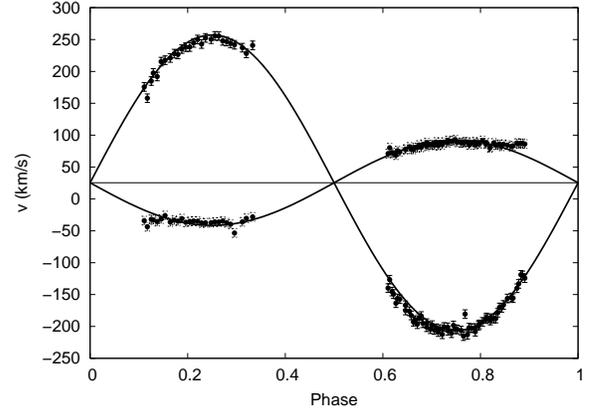}
\end{center}
\caption{Radial velocity curve of DX~Tuc.}
\label{tucrv}
\end{figure}

\begin{figure}
\begin{center}
\leavevmode
\includegraphics[width=8cm]{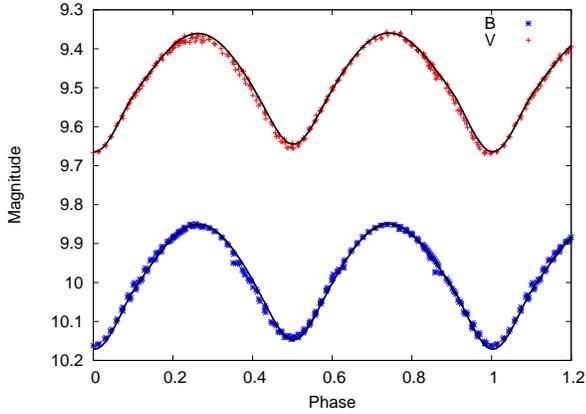}
\end{center}
\caption{Observed and fitted light curves of DX~Tuc.}
\label{tucfit}
\end{figure}

The radial velocity curve and standard photometry  presented here (Fig.\
\ref{tucrv}-\ref{tucfit}) are the first ones in the literature. Using the Hipparcos
ephemeris (HJD$_{\rm min}=2448500.2540$) and the new times of minimum light we
calculated an improved orbital period $P_{\rm orb}=0.37711010(2)$ d. 

The full set of physical parameters is given in Table\ \ref{tucpars}. The system is
a typical contact binary, which has two components of very different
masses but similar temperatures. Nevertheless, the larger star is slightly hotter,
which puts DX~Tuc among the A-type W~UMa systems. The light curve shows a small but
detectable O'Connell-effect ($\Delta V=0.01$ mag), so we included a spot in the light
curve model. With no indication of a third component in the cross-correlation profile,
we fixed $l_3=0$. 
For calculating the absolute parameters, we assumed zero interstellar reddening, which
is supported by the reddening map of Schlegel, Finkbeiner \& Davis (1998) that
implies 
$E(B-V)\leq0.017$ mag in this direction.

\subsection{QY~Hydrae}

The variability of QY~Hydrae was discovered by the Hipparcos satellite (ESA 1997) and
the variable star designation was given by Kazarovets et al. (1999). Although it belongs
to the 100 brightest X-ray stars within 50 parsecs of the Sun (Makarov 2003), no
detailed study has been done so far. (We note that Makarov (2003) listed the star
erroneously as QY~Lyr.) The late spectral type (K2V) and the short orbital period 
put QY~Hya in the XO class of X-ray active stars by Makarov (2003), whose group
includes binary stars of BY~Dra-type, detached binaries (Algols) and eclipsing
binaries of $\beta$~Lyr type. Selam (2004) clearly separated QY~Hya from the W~UMa 
stars using the Fourier-description of the Hipparcos light curve, confirming the 
$\beta$~Lyr class. There is one measurement of the radial velocity, $+25.4\pm0.6$
km~s$^{-1}$, published by Nordstr\"om et al. (2004). They observed the star among
$\sim$14,000 F and G dwarfs, thus avoiding the detection of possible time variability of 
the radial velocity. 

\begin{figure}
\begin{center}
\leavevmode
\includegraphics[width=8cm]{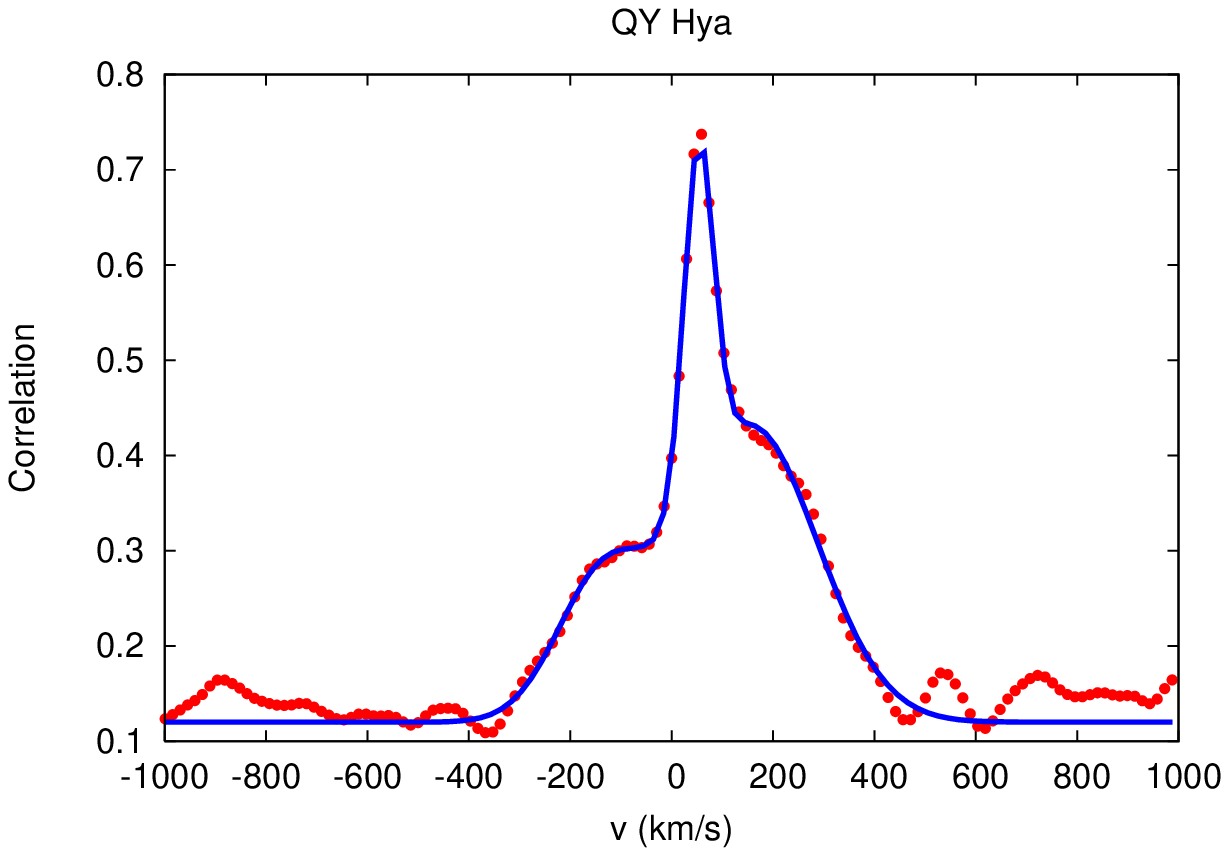}
\includegraphics[width=8cm]{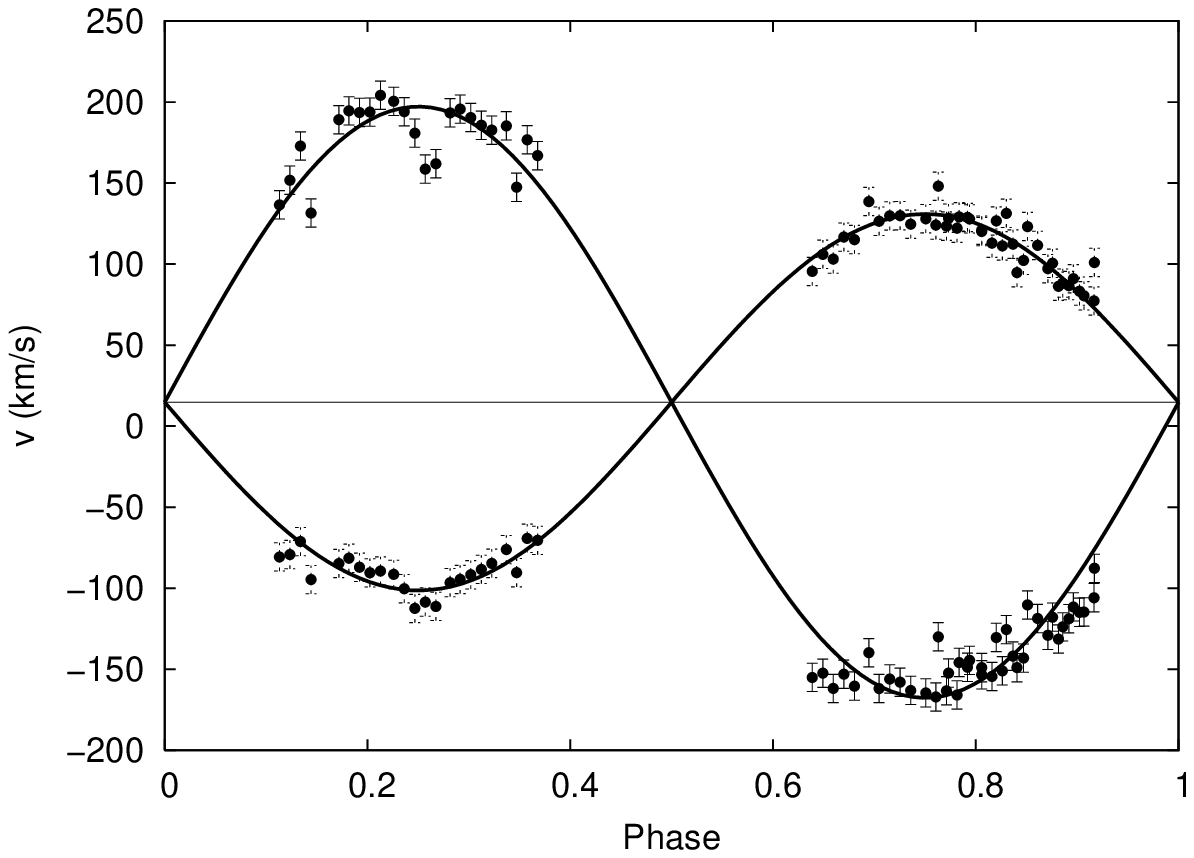}
\end{center}
\caption{{\it Top panel:} the CCF-profile of QY~Hya in phase 0.75 with the 
three Gaussian components. 
{\it Bottom panel:} radial velocities of the eclipsing components of QY~Hya.}
\label{hyaccf}
\end{figure}

\begin{figure}
\begin{center}
\leavevmode
\includegraphics[width=8cm]{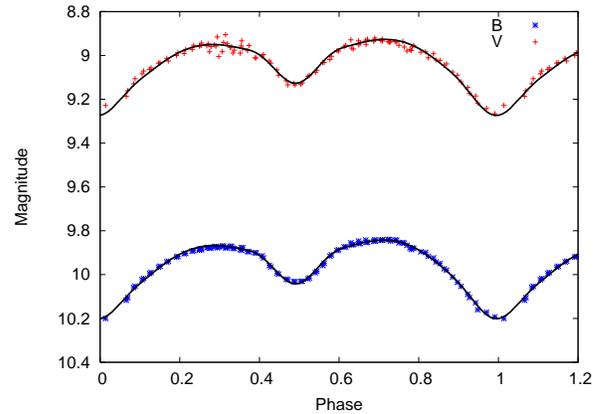}
\end{center}
\caption{Observed and fitted light curves of QY~Hya.}
\label{hyafit}
\end{figure}

\begin{table}
\caption{Physical parameters of QY~Hya.  Underlined values were calculated and fixed by 
the WD-code. HJD$_3$
and $V_3$ refer to mean epoch and radial velocity of the third component.}
\setlength{\tabcolsep}{2mm}
\begin{center}
\begin{tabular}{|llll|}
\hline
Parameter & Value & Parameter & Value\\
\hline
\hline
Spectroscopy & & & \\
$A\sin i$ [R$_{\odot}$] & 1.68$\pm$0.10 & $K_{1}$ [km/s] & 117.0$\pm$8.1\\
$V_{\gamma}$ [km/s] & 14.8$\pm$1.6 & $K_{2}$ [km/s] & 178.1$\pm$8.4\\
HJD$_3$ & 2453186.4 & $V_3$ [km/s] & 53.1$\pm$2.4\\
\hline
Light curve &  fit & & \\
$i$ [$^{\circ}$] & 63.1$\pm$0.6 & $r_1^{\rm side}$ & 0.4128$\pm$0.0034\\
phase shift & $-$0.009$\pm$0.001 & $r_2^{\rm side}$ & 0.3079$\pm$0.0034\\
$q$ &{\it 0.66}($\pm$0.07) & $r_1^{\rm back}$ & 0.4421$\pm$0.0048\\
$T_{1}$ [K] &{\it 5030} & $r_2^{\rm back}$ & 0.3287$\pm$0.0048\\
$T_{2}$ [K] & 5270$\pm$84 & $f$ & ---\\
$\Omega_{1}$ &\underline{3.177} & $(\frac{L_{1}}{L_{1}+L_{2}})_{B}$ & 0.552$\pm$0.012\\
$\Omega_{2}$ & 3.357$\pm$0.044 & $(\frac{L_{1}}{L_{1}+L_{2}})_{V}$ & 0.569$\pm$0.010\\
$r_1^{\rm pole}$ & 0.3907$\pm$0.0028 & $l_{3}$ (B) & 0.207$\pm$0.077\\
$r_2^{\rm pole}$ & 0.2974$\pm$0.0028 & $l_{3}$ (V) & 0.154$\pm$0.041\\
$(B-V)_1$ & 0.91 & $(B-V)_3$ & 0.63 \\
\hline
Spot &  parameters  & &  \\
Co-lat. [deg] & {\it 33} & Rad. [deg] & {\it 46}\\
Long. [deg] & {\it 350} & $T_{\rm fact}$ & 0.84$\pm$0.01\\
\hline
Absolute & parameters & & \\
$M_{1}$ [M$_{\odot}$] & 0.667$\pm$0.014 & $M_{2}$ [M$_{\odot}$] & 0.442$\pm$0.017\\
$R_{1}$ [R$_{\odot}$] & 0.80$\pm$0.03 & $R_{2}$ [R$_{\odot}$] & 0.60$\pm$0.03\\
$L_1$ [L$_\odot$] & 0.37$\pm$0.06 & $L_2$ [L$_\odot$] & 0.25$\pm$0.01\\
$(M_{\rm bol})_1$ & 5.80$\pm$0.13 &  $(M_{\rm bol})_2$ & 6.23$\pm$0.13\\
$(M_{\rm V})_1$ & 6.24$\pm$0.13 &  $(M_{\rm V})_2$ & 6.51$\pm$0.13\\
Sp. type (3) & G4 & $M_3$ [M$_\odot$] & 0.97\\
$d$ [pc] & 50$\pm$2 & & \\
\hline
\end{tabular}
\end{center}
\label{hyapars}
\end{table}

We obtained standard $BV$ light curves with full phase coverage and an excellent
coverage of the radial velocity curve (Figs.\ \ref{hyaccf}-\ref{hyafit}). The
cross-correlation profile shows the presence of a third light at $V_3\approx+50$
km~s$^{-1}$ (top panel in Fig.\ \ref{hyaccf}).  Using the Hipparcos ephemeris
(HJD$_{\rm min}=2448500.2490$) and our new epochs of minimum light, we determined an
updated mean period of $P_{\rm orb}=0.29234050(8)$ d.

The light curve fit indicates a semi-detached binary of similar components (Table\
\ref{hyapars}). There is a slight but significant brightness difference of  $\Delta
V=0.035$ mag between the two maxima, which can be explained by a dark spot on the
primary component. The calculated absolute parameters are based on the assumption of
negligible interstellar reddening (the reddening map of Schlegel, Finkbeiner \& Davis
(1998) gives an upper limit of $E(B-V)\leq0.07$ mag towards QY~Hya, so the $\sim$50
pc distance to the star indeed implies a small colour excess). The third light seems
to have similar contributions in $B$ and $V$, like the third star in VZ~Lib, 
which suggests a hierarchic triple system of three similar components.

\subsection{V870~Arae}

V870~Ara is another Hipparcos discovery, classified as an F8-type contact binary (ESA
1997, Kazarovets et al. 1999). Pribulla, Kreiner \& Tremko (2003) listed the star in
their catalog of field contact binary stars. Selam (2004), based on the
Fourier-decomposition of the Hipparcos light curve, put it among the 64 genuine
W~UMa-type variables, while  Pribulla \& Rucinski (2006) did not find any indication
for multiplicity (caused largely by the lack of data in the literature).

\begin{table}
\caption{Physical parameters of V870~Ara.}
\setlength{\tabcolsep}{2mm}
\begin{center}
\begin{tabular}{|llll|}
\hline
Parameter & Value & Parameter & Value\\
\hline
\hline
Spectroscopy & & & \\
$A\sin i$ [R$_{\odot}$] & 2.43$\pm$0.13 & $K_{1}$ [km/s] & 23.3$\pm$8\\
$V_{\gamma}$ [km/s] & 11.5$\pm$0.8 & $K_{2}$ [km/s] & 283.5$\pm$8.1\\
\hline
Light curve &  fit & &\\
$i$ [$^{\circ}$] & 70.0$\pm$0.5 & $r_1^{\rm side}$ & 0.6424$\pm$0.0007\\
phase shift & 0.0023$\pm$0.0005 & $r_2^{\rm side}$ & 0.2109$\pm$0.0007\\
$q$ &{\it 0.082}($\pm$0.030) & $r_1^{\rm back}$ & 0.6634$\pm$0.0008\\
$T_{1}$ [K] &{\it 5860} & $r_2^{\rm back}$ & 0.3033$\pm$0.0081\\
$T_{2}$ [K] & 6210$\pm$35 & $f$ & 96.4\%\\
$\Omega_{1}$ & 1.849$\pm$0.001 & $(\frac{L_{1}}{L_{1}+L_{2}})_{B}$ & 0.852$\pm$0.001\\
$\Omega_{2}$ & 1.849 & $(\frac{L_{1}}{L_{1}+L_{2}})_{V}$ & 0.860$\pm$0.001\\
$r_1^{\rm pole}$ & 0.5653$\pm$0.0004 & $l_{3}$ (B) &{\it 0.0}\\
$r_2^{\rm pole}$ & 0.1996$\pm$0.0005 & $l_{3}$ (V) &{\it 0.0}\\
\hline
Spot & parameters  & &\\
Co-lat. 1 [deg] & {\it 90} & Rad. 1 [deg] & {\it 17}\\
Long. 1 [deg] & {\it 84} & $T_{\rm fact}$ 1 & 0.90$\pm$0.01\\
Co-lat. 2 [deg] & {\it 57} & Rad. 2 [deg] & {\it 26}\\
Long. 2 [deg] & {\it 0} & $T_{\rm fact}$ 2 & 0.94$\pm$0.01\\
\hline
Absolute & parameters & & \\
$M_{1}$ [M$_{\odot}$] & 1.503$\pm$0.011 & $M_{2}$ [M$_{\odot}$] & 0.123$\pm$0.002  \\
$R_{1}$ [R$_{\odot}$] & 1.67$\pm$0.01 & $R_{2}$ [R$_{\odot}$] & 0.61$\pm$0.01\\
$L_1$ [L$_\odot$] & 2.96$\pm$0.30 & $L_2$ [L$_\odot$] & 0.50$\pm$0.01\\
$(M_{\rm bol})_1$ & 3.54$\pm$0.10 &  $(M_{\rm bol})_2$ & 5.48$\pm$0.10\\
$(M_{\rm V})_1$ & 3.74$\pm$0.10 &  $(M_{\rm V})_2$ & 5.65$\pm$0.10\\
$d$ [pc] & 107$\pm$5 & & \\
\hline
\end{tabular}
\end{center}
\label{arapars}
\end{table}

\begin{figure}
\begin{center}
\leavevmode
\includegraphics[width=8cm]{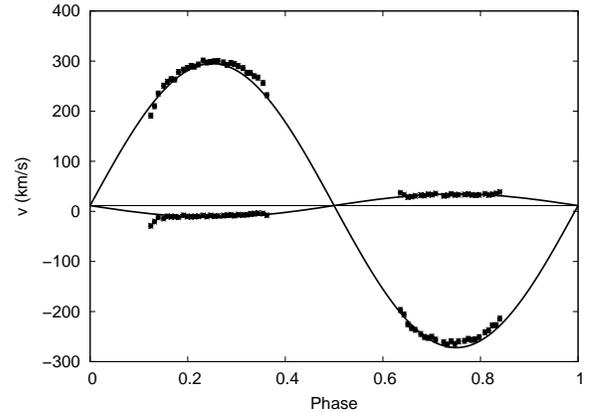}
\end{center}
\caption{Radial velocity curve of V870~Ara.}
\label{ararv}
\end{figure}

\begin{figure}
\begin{center}
\leavevmode
\includegraphics[width=8cm]{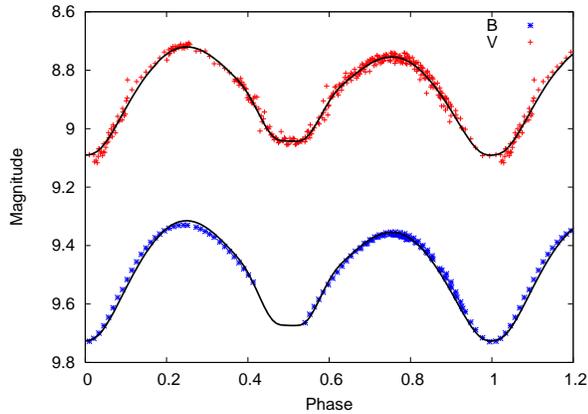}
\end{center}
\caption{Observed and fitted light curves of V870~Ara.}
\label{arafit}
\end{figure}

The radial velocity curve and standard photometry  presented here (Fig.\
\ref{ararv}-\ref{arafit}) are the first ones in the literature. Using the Hipparcos
ephemeris (HJD$_{\rm min}=2448500.1840$) and the new times of minimum light we
calculated an improved orbital period $P_{\rm orb}=0.39972200(2)$ d.

The most interesting feature about V870~Ara is the very small mass-ratio
$q=0.082\pm0.030$. There are  only two contact binaries with spectroscopically
measured mass-ratios around or below 0.08: AW~UMa  ($q\approx0.075$, Rucinski 1992),
SX~Crv ($q\approx0.066$, Rucinski et al. 2001); and one star, V857~Her, for which  the
best-fit light curve solutions strongly suggest a mass-ratio somewhat less than 0.07 but
lacking spectroscopic confirmation (Qian et al. 2005). The existence of these stars are
important, because theory currently predicts a cutoff at about $q=0.09$ (Rasio 1995),
which might be pushed a bit lower to $q=0.076$ (Li \& Zhang 2006). Below that contact
binaries are expected to merge a single fast-rotating star within 10$^3$--10$^4$ yrs.
This puts V870~Ara among the objects that have the potential of constraining evolutionary
scenarios of binary mergers.

Because of the well expressed O'Connell-effect ($\Delta V=0.032$ mag) and asymmetric
distortions of the light curve we added two spots to the light curve solution. 
They rather represent the difficulties we met during the light curve modelling than
two real compact features on the hot component. The finally adopted set of 
parameters (Table\ \ref{arapars}) shows that despite its extreme mass-ratio, this 
is a typical W-type W~UMa system.

\section{Summary}

In this paper we presented new photometric and spectroscopic data and their basic
analysis for five close eclipsing binary stars. The sample consisted of three
southern and two equatorial variables, of which the southern objects have never been
observed and modelled since the discovery. The main results of this investigation can
be summarized as follows:

\begin{itemize}

\item[-] XY~Leo is a hierarchic quadruple system with a W~UMa-type contact 
binary and a BY~Dra-type red dwarf binary. It is one of the best cases for the 
light-time effect in a periodic variable star. Besides determining new spectroscopic
elements and a light curve solution, we also found weak evidence for short-period
magnetic cycles. 

\item[-] VZ~Lib is another multiple system in which we detect the third component both
spectroscopically and photometrically. Recent data indicate the possibility of
detectable light-time effect, thus further eclipse timings are needed to measure the
orbital period of the tertiary companion.

\item[-] DX~Tuc is a typical A-type W~UMa star (i.e. the larger component is hotter).

\item[-] QY~Hya, being one of the 100 brightest X-ray active stars within 50 
parsecs of the Sun, is a late-type triple system with a semi-detached eclipsing pair.  

\item[-] Finally, V870~Ara is a contact binary with the third smallest spectroscopic
mass-ratio in all W~UMa stars to date. 

\end{itemize}

\begin{figure}
\begin{center}
\leavevmode
\includegraphics[bb=50 50 316 301,width=8cm]{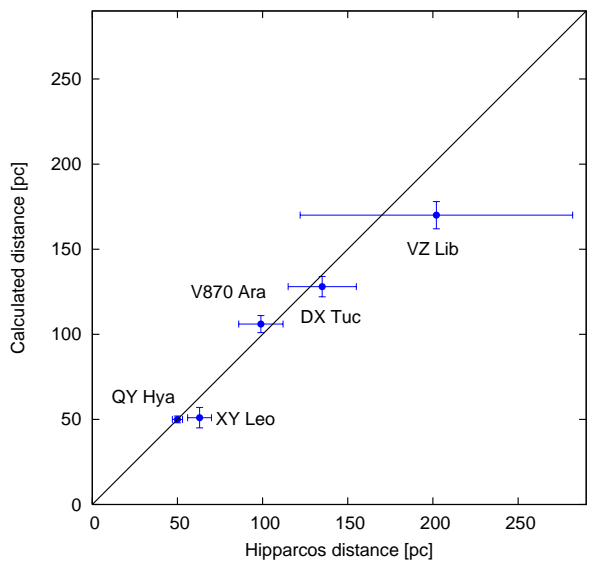}
\end{center}
\caption{A comparison of the calculated distances with the Hipparcos
measurements.}
\label{distance}
\end{figure}

The consistency of the presented results can be tested with the Hipparcos distances.
A comparison of the calculated distances that are based on the light curve models and
the parallax-based Hipparcos values show a good agreement for all the programme stars
(Fig.\ \ref{distance}). In case of DX~Tuc and VZ~Lib, the distances from light curve
models are likely to be improvements over the Hipparcos values. We note that both the
two most deviant stars, XY~Leo and VZ~Lib, have bright tertiary components, which may
have introduced a systematic error in the Hipparcos astrometry that can explain the
larger disagreement.

\section*{Acknowledgments} 

This work has been supported by a University of Sydney Postdoctoral Research
Fellowship, the Australian Research Council and the Hungarian OTKA Grants  \#T042509
and \#TS049872. We thank A. Derekas for assisting the photometric observations in
Siding Spring, Australia. The NASA ADS Abstract Service was used to access data and
references.

\end{document}